\documentclass[twocolumn,times]{aastex701}

\usepackage[T1]{fontenc}
\usepackage{pslatex}
\usepackage{contour}
\usepackage{color}

\usepackage{graphicx}
\usepackage{epsfig} 
\usepackage{natbib}
\usepackage{times}
\usepackage{amssymb,amsmath}
\usepackage{color}
\usepackage{afterpage}
\usepackage{comment}

\definecolor{dark-red}{rgb}{0.8,0,0}
\definecolor{dark-green}{rgb}{0,0.4,0}
\definecolor{dark-blue}{rgb}{0,0,0.8}
\definecolor{dark-magenta}{rgb}{0.8,0,0.8}
\definecolor{orange}{rgb}{1.0,0.6,0}
\definecolor{grey}{rgb}{0.6,0.6,0.6}
\definecolor{gnuplot-purple}{rgb}{.75294117647058823529,.50196078431372549019,1}
\definecolor{darker-purple}{rgb}{0.45176471, 0.30117647, 0.6}
\definecolor{closed}{rgb}{0, .302, .251}
\definecolor{disconnected}{rgb}{.388, .714, 1}
\definecolor{openfolded}{rgb}{1, .757, .027}
\definecolor{opennotfolded}{rgb}{.847, .106, .377}
\definecolor{opentotal}{rgb}{.925, .431, .204}

\newcommand{\stth}[1]{}

\newcommand{\p}[1]{{{#1}}}
 
\shortauthors{Lionello et al.}
\shorttitle{Magnetic Connectivity in the Time-Dependent Corona and
Heliosphere}

\begin{document}

\title{Magnetic Connectivity in the Time-Dependent Corona and Heliosphere}

\author[0000-0001-9231-045X]{Roberto~Lionello}
\affiliation{Predictive Science Inc., 9990 Mesa Rim Rd., Ste. 170, San Diego, CA 92121, USA} 
\email{lionel@predsci.com}

\author[0000-0003-1759-4354]{Cooper~Downs}
\affiliation{Predictive Science Inc., 9990 Mesa Rim Rd., Ste. 170, San Diego, CA 92121, USA} 
\email{cdowns@predsci.com}

\author[0000-0002-8767-7182]{Emily~I.~Mason}
\affiliation{Predictive Science Inc., 9990 Mesa Rim Rd., Ste. 170, San Diego, CA 92121, USA} 
\email{emason@predsci.com}

\author[0000-0003-1662-3328]{Jon~A.~Linker}
\affiliation{Predictive Science Inc., 9990 Mesa Rim Rd., Ste. 170, San Diego, CA 92121, USA} 
\email{linkerj@predsci.com}

\author[0000-0002-1859-456X]{Pete~Riley}
\affiliation{Predictive Science Inc., 9990 Mesa Rim Rd., Ste. 170, San Diego, CA 92121, USA} 
\email{pete@predsci.com}

\author[0000-0003-2061-2453]{Mathew~J.~Owens}
\affiliation{Department of Meteorology, University of Reading, Reading, UK} 
\email{m.j.owens@reading.ac.uk}

%---------------------------------------------------------------------
% ABSTRACT
%---------------------------------------------------------------------
\begin{abstract}
Magnetic flux fills the heliosphere, expands outward from
the solar corona, and is fundamentally related to the structure and dynamics of the solar corona and solar wind.
Open magnetic flux and the fast wind are thought to originate
from open magnetic field lines in coronal holes. Less understood processes in the streamer belt and the boundaries of coronal holes, associated with 
the more variable slow wind, may be formed by interchange reconnection between
open and closed magnetic flux.  Interchange reconnection is thought to give rise to field lines that are ``folded,'' i.e. that turn back on themselves.
The properties of strahl electrons measured in the solar wind give clues to the heliospheric magnetic connectivity.  Unidirectionally outward strahl indicates open field lines, while bidirectional strahl is associated with closed magnetic flux and CMEs.  Inward directed, unidirectional strahl is believed to indicate folded flux.
We use two time-dependent, flux-evolutionary MHD models of the combined corona and heliosphere, one for a solar-minimum configuration, one for the 2024 total solar eclipse, to investigate the magnetic connectivity of the corona/heliosphere system.
We examine how magnetic connectivity varies with distance from the Sun
in the two configurations. We evaluate the evolutionary effects by
contrasting time-dependent results  with the corresponding steady-state calculations, and compare the model connectivities with statistical studies of strahl.  The connectivities in the time-evolving simulations are roughly consistent with observed strahl occurrence rates, while those from the steady-state models are not.  Our results suggest that complex magnetic connectivities are ubiquitous in the heliosphere.

\end{abstract}


%---------------------------------------------------------------------
% SECTION: INTRODUCTION
%---------------------------------------------------------------------
\section{Introduction}
\label{s:intro}
The solar magnetic field expands outward from the solar surface and fills the corona and the heliosphere.  It plays a key role in the physics of the solar wind.
Many of the outstanding questions in heliophysics revolve around understanding how the properties that are measured in situ in the solar wind were created back at the Sun.  In the presence of ideal  flows and in the reference frame co-rotating with the Sun, the solar wind plasma flow is aligned with the magnetic field. In this approximation, tracing the magnetic connectivity of plasma parcels encountered in the heliosphere back to the Sun reveals their solar origin. The magnetic field connectivity is also important for the transport and propagation of solar energetic particles (SEPs), which are guided along magnetic field lines from their generation near the Sun to locations in the heliosphere.

In terms of solar connectivity, magnetic field lines can either be open (one end
connected with the photosphere), closed (both ends connected with the
photosphere), or disconnected (neither end connected with the photosphere).
In the standard paradigm of coronal structure \citep[e.g.,][]{mackay_yeates2012,priest2014,linkeretal2017}, the open field originates
primarily in coronal holes (CHs), which are defined as low-intensity emission regions in EUV and in X-rays  \citep{1977chhs.conf...27B,1977RvGSP..15..257Z}, while magnetically closed field regions form the streamer belt.  From the low coronal perspective, open field lines are simply those that stretch far enough away from the Sun that they are carried out by the solar wind and no longer confine the plasma near the Sun.  This definition is ambiguous from the heliospheric perspective, as these different connectivities may be present in the  field encountered by in situ spacecraft.  The properties of the field in the heliosphere appear to be associated with the origin of the solar wind, where fast wind is found to emanate from deep within coronal holes and expected to be ``open'' in the heliosphere.  The slow wind is associated with the streamer belt and the boundaries of CHs \citep{2016SSRv..201...55A}.   \citet{1998SSRv...86...51F} argued that so-called interchange reconnection between  closed and open magnetic flux releases  the plasma that forms the slow solar wind.
The topological properties associated with this process are elucidated in
the S-web model \citep{2011ApJ...731..112A,2011ApJ...731..110L,2011ApJ...731..111T}.   Interchange reconnection within the S-web is also expected to create magnetic fields that are locally inverted or turned back on themselves \citep{crookeretal2004}, that we refer to here as ``folded'' flux. 
Folded flux could also be generated by shear in adjacent solar wind flux tubes which causes the folds to develop over time in the heliosphere as argued by \citep{2018ApJ...868L..14O}. Such shear patterns could also result from interchange reconnection, and measurements of $B_r$ \citep{2009JGRA..11411104L} and of
strahl at a range of heliocentric distances \citep{2020MNRAS.494.3642M} support this possibility.

While the true connectivity of the magnetic field lines cannot be directly measured in situ, 
 the properties of electrons propagating in the solar wind can provide important clues about their connectivity and geometry.
Results from the HELIOS plasma experiment showed that  electrons usually
exhibit a suprathermal tail known as strahl \citep{1977JGZG...42..561R}.
When a strahl is observed,
 it is often in one direction, i.e.\ going outward from the Sun either parallel or anti-parallel to the heliospheric magnetic field. However,
counterstreaming strahls are also present, which may be evidence of
closed magnetic field lines or, as \citet{1987JGR....92.8519G} argued,
magnetic flux  connected with coronal mass ejections (CMEs).
\citet{2012JGRA..117.4107A}  presented
 a statistical survey of strahls using measurements from 1998 to 2002 and
found that a strahl is present $\geq 75\%$ of the time, with a unidirectional strahl present $65\%$ of the time, and a 
 $10\%$ occurrence rate for counterstreaming electrons.   \citet{2017JGRA..12210980O} noted the occurrence of unidirectional, sunward directed strahl as an indication of folded magnetic flux, and emphasized its importance for estimating the open magnetic flux in the heliosphere, which can be overestimated if simple averaging of the the interplanetary unsigned flux is employed \citep{owensetal2008}.  \citet{2017JGRA..12210980O}  provided estimates of the occurrence rates of all four strahl:   unipolar outward ($\sim$69\%), bidirectional ($\sim$4\%), inverted ($\sim$17\%), and absent ($\sim$10\%).  \citet{2022SoPh..297...82F} expanded upon the  \citet{2017JGRA..12210980O} study to look at statistics over 27 years, finding similar occurrence rates, although notably more bidirectional strahl ($\sim$11\%). These differences result from thresholds used to define the presence of strahl at 0 and 180 degrees.
 
In this work we explore 
how magnetic connectivity varies temporally and spatially
 in the solar corona and heliosphere and find  
 that this is an essentially dynamic process. To show this, we use
time-dependent,
flux-evolutionary simulations of the corona \citep{2023ApJ...959...77L, 2025Sci...388.1306D} and heliosphere,
with the heliospheric evolution coupled to the corona
using the method of \citet{2013ApJ...777...76L}.
To understand the importance of flux evolution, time-dependent
results are compared with the correspondent steady-state cases
 \citep{2023ApJ...959L...4M}.
 Two different
configurations are  considered here: a solar-minimum case \citep[as in][]{2023ApJ...959...77L} and the time around the 2024 total solar
eclipse, near solar maximum \citep{2025Sci...388.1306D}. For each configuration,
 we calculate the magnetic connectivity as a
function of the distance from the solar surface
 and determine 
 what fractions of open, closed, and disconnected
flux are present at 1 AU.  We compare the model results with strahl occurrence rates from the \citet{2012JGRA..117.4107A}, \citet{2017JGRA..12210980O}, and \citet{2022SoPh..297...82F} studies.

 In \S~\ref{sec-model}
we describe the combined models and the computations; in \S~\ref{sec-connect}
we show and compare connectivity in the models. In the last section
we discuss the physical implications of our results.

%---------------------------------------------------------------------
% SECTION: SIMULATIONS
%---------------------------------------------------------------------
\section{THE COMBINED, TIME-DEPENDENT MHD MODELS}
\label{sec-model}

%---------------------------------------------------------------------
% TABLE: SIMULATION LIST
%---------------------------------------------------------------------
\begin{deluxetable*}{ccccccc}
\tablewidth{0pt}
\tablecaption{Computer simulations considered in this work
\label{tab:runs}}
\tablehead{
\colhead{Label} \vspace{-0.2cm}   &  \colhead{Coronal} & \colhead{Heliospheric} &
\colhead{Outer} &
\colhead{Conditions} & \colhead{Flux}& \colhead{References}\\ 
 & \colhead{Resolution} & \colhead{Resolution} & \colhead{Boundary} &  &
\colhead{Evolution} & for Coronal Run }
\startdata
SMTD & $269\times181\times361$  & $763\times181\times361$ & 430 $R_\odot$  & Solar Minimum &
719~h& \citet{2023ApJ...959...77L} \\
SMSS & $269\times181\times361$  & $401\times181\times361$ & 230 $R_\odot$ & Solar Minimum &
 No & \citet{2023ApJ...959L...4M} \\
E24TD & $284\times292\times630$  & $1598\times334\times720$ & 430 $R_\odot$ &  Eclipse 2024 &
775~h & \citet{2025Sci...388.1306D}\\
E24SS & $284\times292\times630$  & $801\times334\times720$ & 230 $R_\odot$ & Eclipse 2024 & No 
& ---\textquotedbl---
\enddata
\end{deluxetable*}
 
Our calculations have been obtained with the MAS (Magnetohydrodynamic 
Algorithm outside a Sphere) MHD model, which has been used  for modeling
both the solar corona and the heliosphere
\citep[most recently and
respectively,][]{2025Sci...388.1306D,2025ApJ...979..204R}.  
\subsection{The Coronal Model}
The coronal model 
\citep[see][for a full description]{2018NatAs...2..913M,2023ApJ...959...77L}, is a 3D, resistive,
thermodynamic
MHD rendition of the solar corona spanning from the solar surface
to 30 $R_\odot$. A Wave-Turbulence-Driven (WTD) heating and
acceleration is prescribed 
\citep{2014ApJ...784..120L,2016ApJ...832..180D,2018NatAs...2..913M}.
Although we have always advanced MAS  equations in time, in the past we 
used to look for a steady state (SS). Now we also 
follow the time-dependent (TD) evolution of the corona by
incorporating surface flux-evolution at the photospheric boundary to drive the system. This is done by 
prescribing 
at the inner boundary,  $r=R_\odot$, a sequence of magnetic
flux maps as well as 
differential rotation and meridional flows \citep{2023ApJ...959...77L}.
Energization of the magnetic field can also be introduced to obtain 
more realistic structures and better match the observed corona
\citep{2025Sci...388.1306D}.
\subsection{The Heliospheric Model}
From a shell at $r=25~R_\odot$ (i.e., where the
flow is already
 supersonic and super-Alfv\'enic) within the coronal model, we  extract
a time sequence of   fields (magnetic, velocity, density, temperature).
These are  prescribed as boundary 
conditions at the lower boundary of the heliospheric model to
drive its evolution  \citep{2013ApJ...777...76L}.
The outer boundary is  
set at  $r=230$ for steady-state runs, or $430~R_\odot$ for evolutionary runs, a further
 extension  being necessary to minimize boundary effects at 1 AU.
The heliospheric 
model 
employs the so-called polytropic approximation 
\citep{2003PhPl...10.1971L}, which uses
a simple adiabatic energy equation, and, in the present case,
the ratio of specific heats is $\gamma=3/2$.
\subsection{Simulations}
We performed four  simulations using the coupled
coronal and heliospheric models.
These are summarized in  
Table~\ref{tab:runs}. The coronal part of the SMTD simulation
is described in \citet{2023ApJ...959...77L}. The TD
evolution of the corona is calculated
 for $719~h$,  in response to surface magnetic
field  maps and  flows corresponding to a solar minimum (SM) configuration
 obtained from the model of
\citet{2003SoPh..212..165S}. 
The initial condition for the
heliosphere is a potential field extrapolation of the coronal magnetic flux
at  $25~R_\odot$. The same radial  velocity and temperature from
the said surface are prescribed
for each internal point of the same latitude and longitude, while
the density is scaled with $r^{-2}$ \citep{2013ApJ...777...76L}. 
The coronal sequence is then used to drive the heliosphere. 
 Since the signal from the
 driving propagates from the
lower boundary at $25~R_\odot$ outward, it takes
$\approx 100~\mathrm{h}$ for it to reach Earth. During this relaxation time,
conditions at 1 AU only feel the effect of rotation.

For the SMSS simulation, we use  {one of the coronal SS  solutions
presented in \citep{2023ApJ...959L...4M}. It was} 
obtained
by using the surface flux-distribution from the corresponding TD model at $526~\mathrm{h}$ as a static boundary condition and relaxing for about 80~h.  Then the conditions at $25~R_\odot$ from this
relaxed state are 
propagated outward into the heliosphere for about 240~h to obtain a SS throughout.

The E24TD simulation is based on the prediction  of           the
2024 total solar eclipse (E24) as described in \citet{2025Sci...388.1306D}.
For $775~\mathrm{h}$,  photospheric magnetic field observations were 
updated, processed, and
assimilated as boundary conditions at $r=R_\odot$ in near real time \citep{2025ApJS..278...24C}.
Energizing, time-dependent electric fields were also applied 
at the boundary \citep{2018NatAs...2..913M}. The coronal 
calculation is then used to obtain a TD solution for the heliosphere. 
The initialization of the heliosphere is specified  as in the SMTD case.
Then
the values at $25~R_\odot$ extracted from the coronal calculation are used
to drive the heliosphere
for  $775~\mathrm{h}$.
Finally, similar to the SMSS case, the E24SS simulation is obtained
 {from one of the steady-state calculations studied in \citet{2025Sci...388.1306D}, who fixed the surface flux distribution  at $320~\mathrm{h}$ and relaxed the corona for 80~h.} This solution 
is then propagated outward with
the heliospheric MAS for 240~h to obtain an overall SS.

%---------------------------------------------------------------------
% SECTION: CONNECTIVITY
%---------------------------------------------------------------------
\section{CONNECTIVITY}
\label{sec-connect}

\begin{figure}
\includegraphics[width=0.45\textwidth]{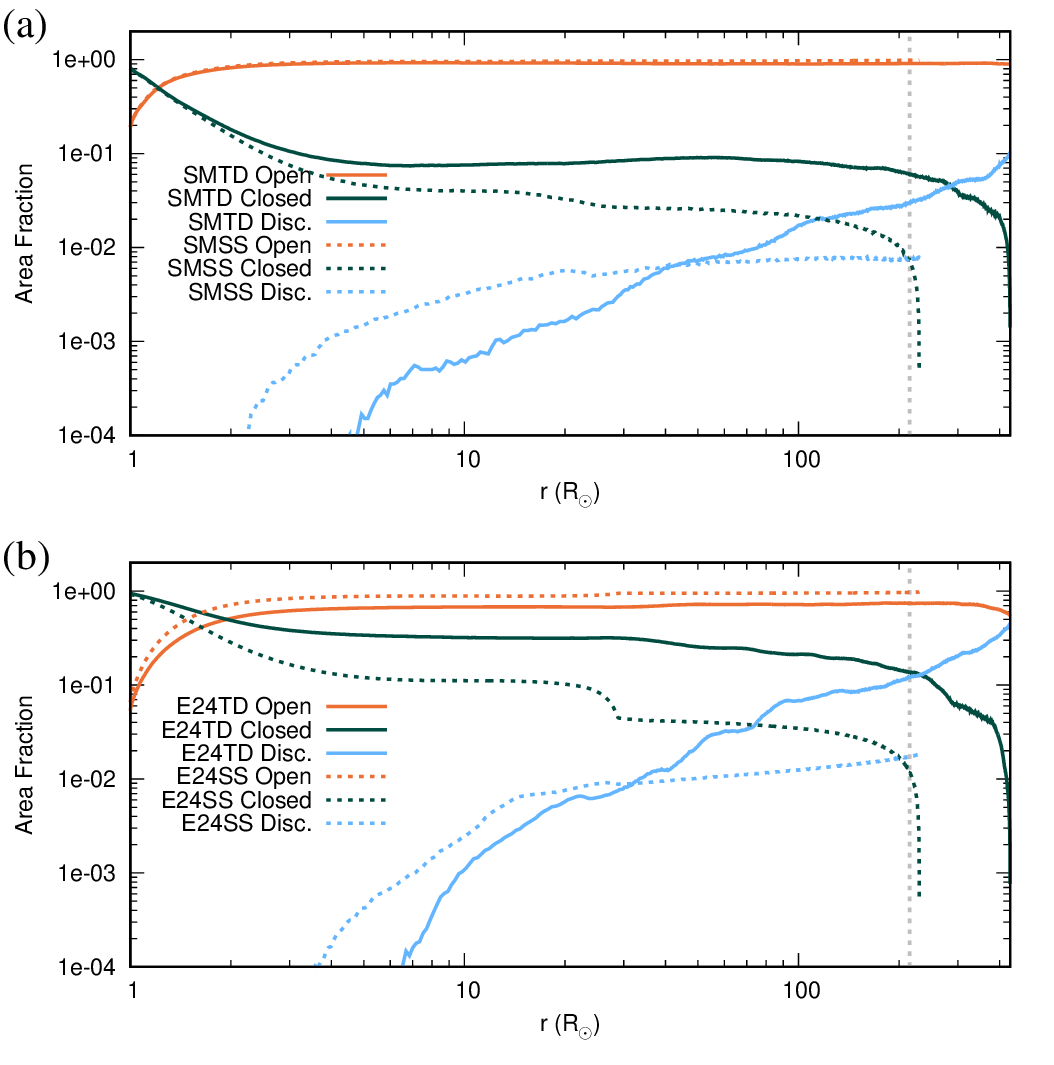}
\caption {Area fractions of the spherical surface at
different radii occupied by magnetic field lines of different
connectivity: (a) SM runs; (b) E24 runs. The dotted vertical line is
1 AU.}
\label{fig-radial}
\end{figure}

We now examine the connectivity by  tracing
 field lines in either direction                
from points within the combined coronal and heliospheric domain. 
Closed regions have both endpoints at $r=R_\odot$, open regions have
one point at  $r=R_\odot$ and one at  the outer boundary, disconnected
regions have both endpoints at  the outer boundary.
For the TD runs,
we trace at times well after the initial transient phase ($\approx
200~\mathrm{h}$),
during which the signal of the first time sequence travels 
from the $25~R_\odot$ shell
to reach the outer boundary at 2~AU.
Figure~\ref{fig-radial} shows the area fractions of spherical surfaces
as a function of radius. The open field fractions are in orange, the
closed field  in green, and the disconnected in sky. The TD quantities
are traced with continuous lines, the SS with dotted lines.
 The (a) panel
is for the SM cases, the (b) panel for the E24.  As one moves outwards, 
the area of open magnetic flux (orange) increases rapidly to fill the corona and 
then  asymptotically tends to a plateau. 
However,   
while SMTD and SMSS have similar open-area
%flux 
percentages, E24TD has distinctively
less open area than E24SS. Unsurprisingly,
 the area of closed magnetic flux (green) decreases 
rapidly in the first few solar radii from the solar surface. SS runs have more open area than TD runs. This is particularly true for E24TD in comparison with E24SS.

%---------------------------------------------------------------------
% FIGURE: AREA FRACTIONS VS. RADIUS
%---------------------------------------------------------------------

The area fraction of disconnected flux (sky blue) remains negligible for the SS simulations ($\lesssim 1\%$) but generally grows with increasing distance in the TD simulations. This difference is related to the inherent dynamical evolution present in the TD simulations, in which the heliospheric current sheet (HCS) periodically reconnects and ejects material and magnetic flux ropes \citep[e.g.,][]{reville20a}. This process brings together the flux immediately adjacent to the HCS, which is generally closed initially but eventually involves open flux on either side. When two open flux tubes on either side reconnect, this forms disconnected flux that maps from the location of the flux-rope as it propagates to the outer boundary. The superposition of several of these structures in height all advecting outward as they leave the domain leads to the relative increase of disconnected flux with height. 
The relative amount of closed and disconnected flux is also increased in the E24TD simulation by the energization scheme incorporated into the boundary driving, which leads to the formation of several flux-ropes and small eruptions originating from the low corona.

Nearing the outer boundary of the TD simulations we see a drop in the closed flux and a rise in the disconnected flux. Because the simulation must include an outer boundary, its presence introduces the potential for miscounting closed flux within flux-ropes as disconnected as the structure nears the outer boundary. This is primarily our motivation for placing outer boundary of the heliospheric simulation at 2~AU while focusing on flux-fractions at 1~AU.

%---------------------------------------------------------------------
% FIGURE: 2D CONNECTIVITY MAPS
%---------------------------------------------------------------------
\begin{figure}
\includegraphics[width=0.45\textwidth]{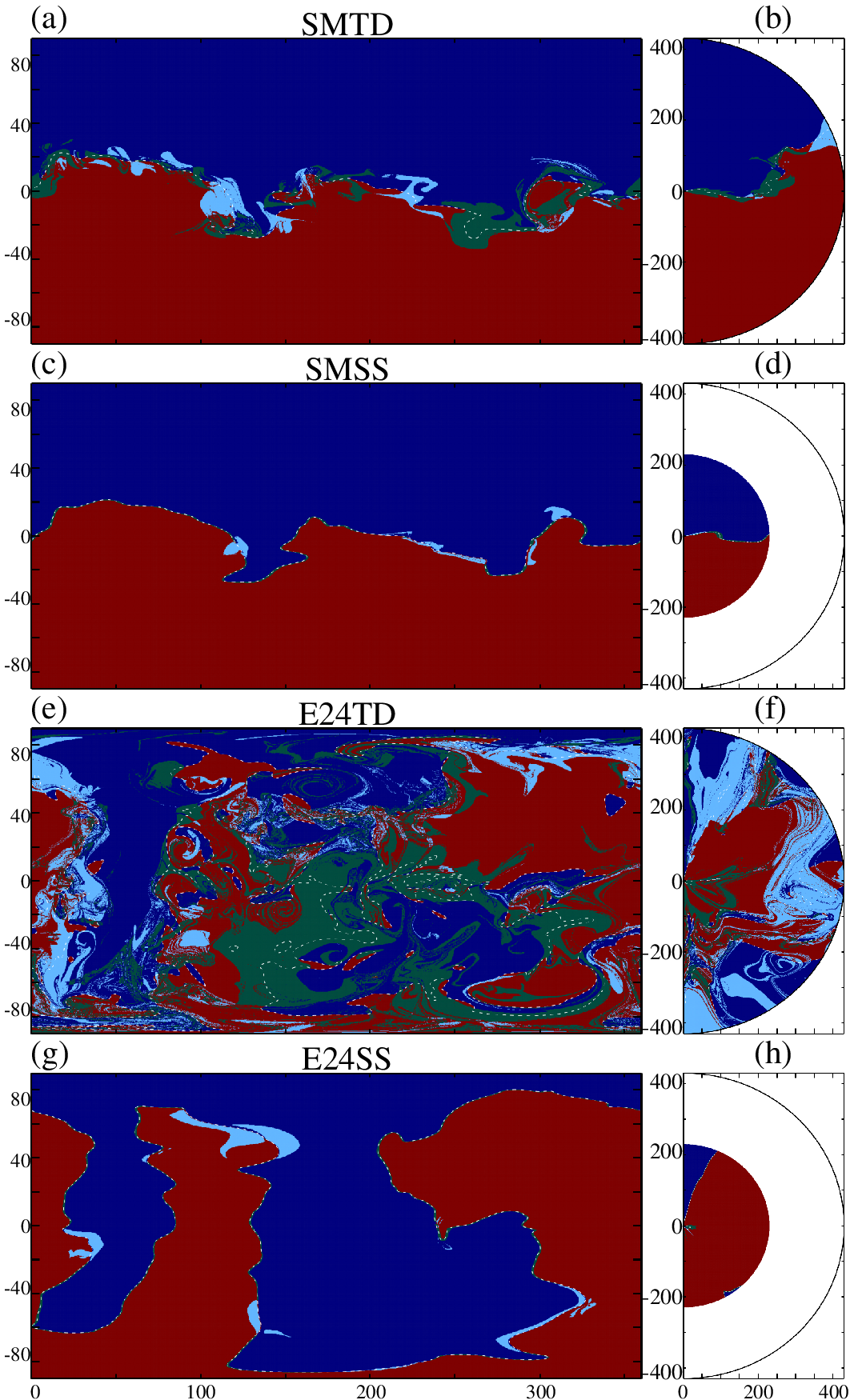}
\caption {Connectivity maps for the runs of Table~\ref{tab:runs}.
Color codes: navy, open negative flux; maroon, open positive
flux; green, closed flux; sky, disconnected flux;  {white dashed} line, $B_r=0$.
\textbf{(a)} Mercator projection 
 at 1 AU  for  SMTD
\textbf{(b)} Cut at $0^\circ$ longitude for SMTD. 
\textbf{(c)}  Same as (a) for
SMSS. \textbf{(d)}  Same as (b) for SMSS.
\textbf{(e)} and \textbf{(f)}:  respectively
the same as (a), (b) for E24TD.
\textbf{(g)} and \textbf{(h)}:  respectively
the same as (c) and (d) for E24SS.
}
\label{fig-maps}
\end{figure}

In Fig.~\ref{fig-maps} we present maps of the connectivity of the runs
of Table~\ref{tab:runs} on Mercator projections at 1 AU (a, c, e, and g)
and in a cut at $\mathrm{longitude} = 0^\circ$ (b, d, f, and h). 
We distinguish between 
open flux of positive  (maroon) and negative (navy) polarity. Sky is 
used for disconnected flux, green for closed flux, and the  {white dashed} line
indicates where $B_r=0$, which for SS runs marks the HCS. In agreement with Fig.~\ref{fig-radial} 
the TD runs show far wider areas of disconnected flux compared
with SS runs. This is especially true for E24TD, (e) and (f), when
compared with E24SS, (g) and (h). Areas of closed flux are also
more extended in the TD runs. The boundaries between 
the different areas are more complex in TD runs than in SS runs. 
This is especially evident if one compares panel (e)  with (g) (E24TD 
vs.\ E24SS).

%---------------------------------------------------------------------
% FIGURE: CONNECTIVITY VS. TIME
%---------------------------------------------------------------------
\begin{figure}
\includegraphics[width=0.45\textwidth]{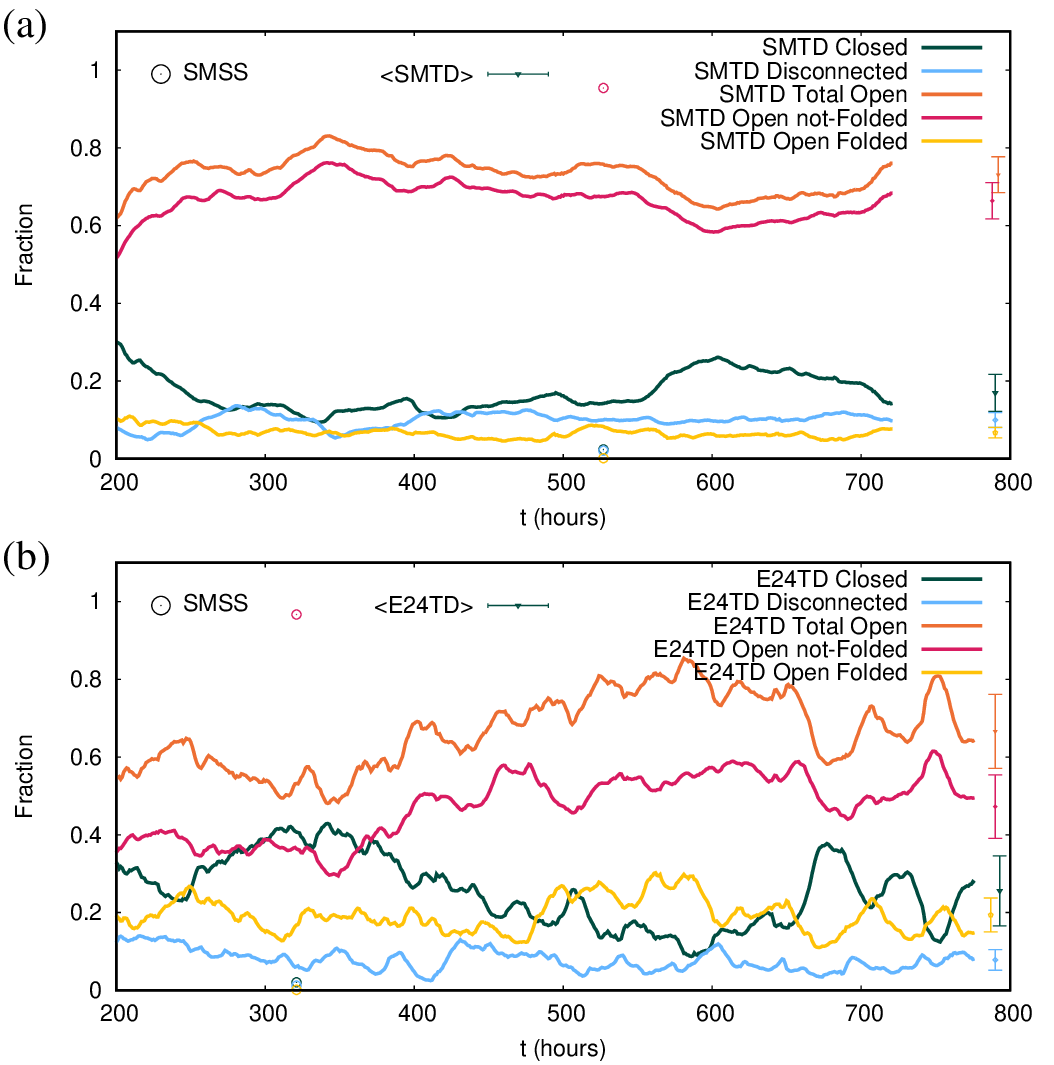}
\caption {Connectivity fractions as a function of time in the 
$\pm 6^\circ$ latitude zone: 
(a) SM runs; (b) E24 runs. We have removed the first 200~h hours
of the simulations, during which heliospheric relaxation phase
occurs.
 Closed flux is green, disconnected is sky, open folded flux
gold, open not-folded rose, total open flux orange. Average values
with standard deviation errors from 200~h to the end of the TD simulations
are shown on the right within each plot. The values for the SS simulations
are shown as circles at 526~h for SMSS and 320~h for E24SS.}
\label{fig-temporal}
\end{figure}

We now turn our attention to the temporal evolution of 
connectivity in the $\pm 6^\circ$  latitude zone at 1 AU. This roughly
corresponds
to the extension of the solar B angle  {(the heliographic latitude of the central point of the solar disk)}, 
 within which the Earth and most spacecraft orbit. 
 The results are displayed in Fig.~\ref{fig-temporal}.
 For each  of the
 simulations of Table~\ref{tab:runs}, we show the area
 fractions of the zone corresponding to each connectivity, $i$. We distinguish between
 folded and not-folded open field
 lines: a point of an open field line
 is considered  folded if it belongs to the sunward portion of the said line.
 Each area fraction $a_i$
 corresponding to each connectivity $i$
 (closed, disconnected, open folded, open not-folded) is
 defined as follows:
 \begin{eqnarray}
     a_i&=&\frac{1}{a}\int_0^\pi \int_0^{2\pi} z(\theta)c_i(\theta,\phi)\sin \theta \, d\theta \, d\phi, \\
     a&=&\int_0^\pi \int_0^{2\pi} z(\theta)\sin \theta \, d\theta \, d\phi, \nonumber
 \end{eqnarray}
 where $z(\theta)$ is 1 in the $\pm 6^\circ$ equatorial zone and 0 elsewhere,
 and $c_i(\theta,\phi)$ is 
 1 where the connectivity is $i$ and 0 elsewhere.
 While the 
 TD simulations have evolving values in time (lines), the SS have
 constant fractions (circles) calculated at the end of the relaxation phase. 
All the flux-fraction lines of SMTD show lower variability than those
of E24TD. This can be appreciated by examining the right section of Table~\ref{tab-comp}, which 
shows the
 average and standard deviation for each connectivity of the TD runs in the 
 interval between 200~h (i.e., after the transient phase) and the end of
 the simulation.  If we compare TD with SS runs, we find that SS runs 
 are dominated by open not-folded flux, while TD runs have
significant fractions of all connectivities. 

To better illustrate the properties of the simulation,
we have flown the trajectory of the Earth in  E24TD from 200~h to the end
of the calculation. We have calculated magnetic field line along the
trajectory and colored them with the same color scheme of Fig.~\ref{fig-temporal}, except that we do not mark the sign of the magnetic flux. The
result is displayed in Fig.~\ref{fig-fl}, which has a 3D rendering
of the orbit on top and a Mercator projection in the bottom. The initial
point is marked with a blue arrow.

%---------------------------------------------------------------------
% TABLE: STRAHL VS. MODEL
%---------------------------------------------------------------------
\begin{table*}
\begin{center}
\begin{tabular}{ |l r r r||r r r r l| }
\hline
Strahl &   A12& O17      &  F22 &SMTD & SMSS & E24TD & E24SS & Connectivity  \\
\hline
Unidirectional	& $\geq 65$ & $85.7$ & $80.5$ &  $73 \pm5$ & 96& $66\pm10$ & 97 & \textcolor{opentotal}{Total Open}   \\
\hspace{0.5em}$\hookrightarrow$ Anti-Sunward &  & $68.7$ 	& $65.5$ & $66 \pm5$ & 96& $47\pm8$ & 97 & \hspace{0.5em}$\hookrightarrow$ \textcolor{opennotfolded}{Open not Folded}   \\
\hspace{0.5em}$\hookrightarrow$ Sunward&  & $17.0$	& $ 15.0$ & $7 \pm1$ & 0& $19\pm4$ & 0 & \hspace{0.5em}$\hookrightarrow$ \textcolor{openfolded}{Open  Folded}   \\
Counterstreaming& $\approx 10$ &$4.1$   & $11.3$ & $17\pm5$  & 2 & $26\pm9$ & 1 & \textcolor{closed}{Closed} \\
Unclassified/Absent& $\leq 25   $ & $10.1$		& $8.3   $ & $10\pm2$ & 2 & $8 \pm 3 $ & 2 & \textcolor{disconnected}{Disconnected} \\
\hline
\end{tabular}
\end{center}
\caption{The strahl statistics of \citet{2012JGRA..117.4107A} (A12),
\citet{2017JGRA..12210980O} (O17), and \citet{2022SoPh..297...82F} (F22) compared with the connectivities of Fig.~\ref{fig-temporal}.
}
\label{tab-comp}
\end{table*}

%---------------------------------------------------------------------
% SECTION: DISCUSSION
%---------------------------------------------------------------------
\section{DISCUSSION AND CONCLUSION}
\label{sec-dis}

We have examined the magnetic connectivity in the corona and heliosphere
in 4 MHD simulations, covering the possible combinations between flux evolution vs.\ steady state, solar minimum vs.\ maximum. As we move outwards from the Sun, the fraction of disconnected flux and folded flux in TD runs grows much faster than
in SS. This can be interpreted as an effect of the photospheric
driving that continuously stresses the corona, leading to reconnection in the HCS and small CMEs
The decrease in closed flux areas is somewhat slower in TD runs
(especially for E24TD) than in SS. These trends lead to dramatically different
pictures at 1 AU, where for SS runs we simply find two areas of open flux 
of opposite polarity separated by the HCS, 
while in TD simulations intertwined areas of disconnected and closed flux  are disseminated along 
the $B_r=0$ curve (or curves) of the HCS. 
We have then focused our attention on the zone at 1 AU where Earth and most
spacecraft orbit, and found quantitative evidence that the connectivity
frequencies of closed, disconnected and (to a lesser degree) open flux between TD and
SS runs are incompatible. 

The connectivity of magnetic field lines appears to be
related to the propagation of electrons and, in particular, with
their suprathermal tail, or strahl. The statistics of long-term studies of strahl observations at 1AU from the  ACE/WIND spacecraft are provided in the right section of Table~\ref{tab-comp} and can be summarized as follows:
The survey of \citet[][hereafter A12]{2012JGRA..117.4107A},
for the 1998--2002 time period (increasing phase towards maximum)
concludes that strahl occurred
$\geq 75 \%$ of the time, while counterstreaming strahls  about $10\%$ of the time. We deduce that 
unidirectional strahl was observed for $\geq 65 \%$ of the
interval and no strahl for
$\leq 25 \%$. In the statitics provided by
\citet[][hereafter O17]{2017JGRA..12210980O} and \citet[][hereafter F22]{2022SoPh..297...82F} a 
further distinction is introduced beside undetermined and 
counterstreaming strahl: unidirectional strahl can either be directed
 away from or towards  the Sun.   This inverse strahl is associated with
a switchback or folding of the magnetic field.

Comparing the observed strahl statistics (Table~\ref{tab-comp}, left) to the average connectivity fractions of our simulations (Table~\ref{tab-comp}, right), we find that the TD and SS simulations contrast quite differently with the observations. Regarding the total open magnetic flux, the SS simulations, by virtue of being almost entirely open ($\ge96\%$), 
largely over-predict the measured frequency of unidirectional strahl. Conversely, the TD simulations under-predict the average frequencies determined from the long-term O17 and F22 surveys but are compatible with the A12 survey, which covers a slightly more active than average period during the ascending phase of solar cycle 23.

Splitting the strahl and open flux statistics into the anti-sunward (open not-folded) and the sunward categories (open and folded) we find that the SS simulations are wholly incompatible with the relative frequency of folded flux ($0\%$ vs. $15$-$17\%$), which is not surprising because such structures cannot typically be produced in a steady state configuration. The TD simulations on the other hand, average between $7$ and $19\%$ folded flux respectively, similar to the observational determinations from the O17 and F22 surveys. This underscores the importance of time-dependent dynamics in forming the folded magnetic structures encountered in the heliosphere at 1AU.

A similar picture emerges when comparing the counterstreaming (closed) and unclassified/absent (disconnected) categories to the SS and TD simulations.  The SS simulations produce very little of either closed or disconnected flux while the TD simulations create a healthy amount, of order $\sim20\!\%$ and $\sim\!10\%$ for each respective category. As with the folded open flux category, the relative fractions from the TD simulations appear much more compatible with the observational studies than the SS simulations, again implying that the SS simulations are missing an essential ingredient to forming these magnetic structures.

\begin{figure*}[h]
\begin{center}
\begin{interactive}{animation}{owensangle2AU_RD.mov}
\includegraphics[width=0.84\textwidth]{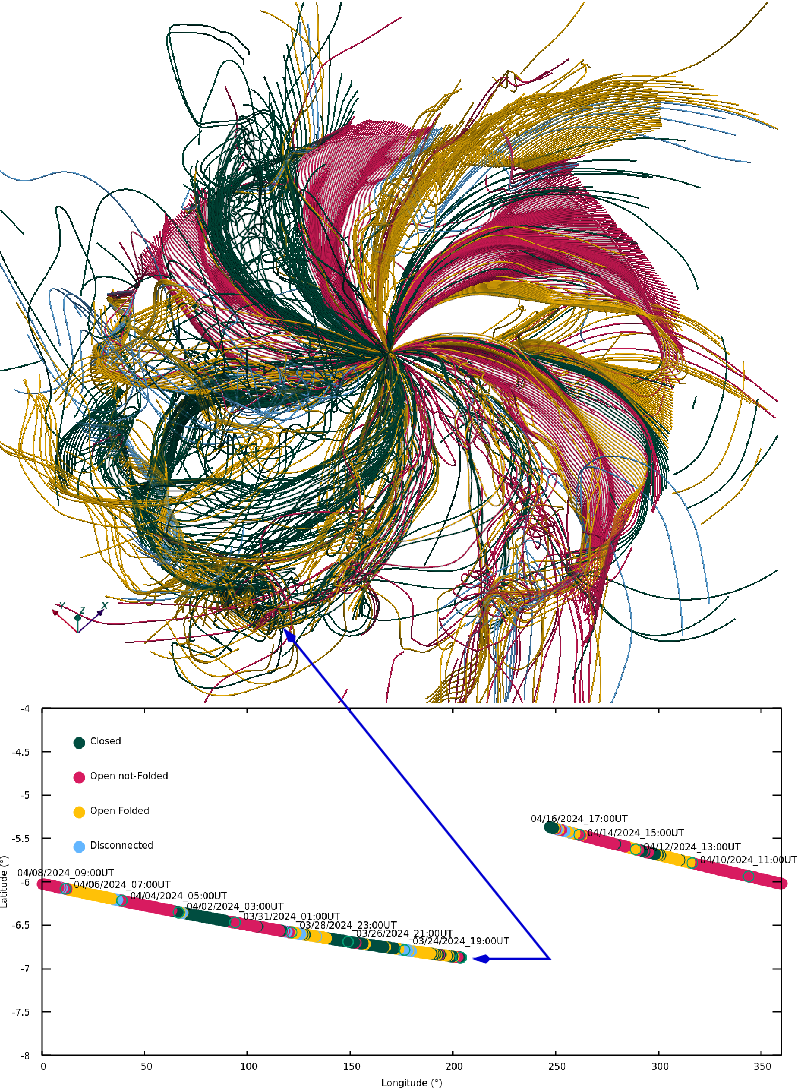}
\end{interactive}
\caption {Field line connectivity along the
trajectory of Earth in the E24TD calculation, from March 24, 2024 at 19:00 UT
(200~h in the simulation) to the end (April 16, 2024, 17:00 UT).
Top: 3D rendering. An animation showing all angles is available in the HTML version. Bottom: the trajectory in the longitude-latitude plane.
As in Fig.~\ref{fig-temporal}, closed field lines are green, open not folded
  rose, open and folded gold,
and disconnected  sky. The double blue arrow  shows the beginning
of the trajectory.
}
\label{fig-fl}
\end{center}
\end{figure*}

That said, caution must be exercised when explicitly ascribing observed strahl signatures to different connectivity categories of traced magnetic field lines.  As discussed in A12, electron scattering processes during transit along magnetic field lines can influence the relative width and amplitude of the suprathermal electron population. The importance of scattering is highly correlated with the overall field-line length (fade-out) as well as the local plasma properties encountered, both of which can vary significantly between `simple' spiral flux-tubes in the heliosphere and the folded or twisted magnetic field lines within the HCS or ICMEs (Fig.~\ref{fig-fl}). Additionally, the observational criteria used to delineate different connectivity categories based on strahl is not uniform between studies, and has led to variation in the measured statistics.

For example, comparing the relative frequency of closed flux in the TD simulations ($\sim20\!\%$) to the counterstreaming frequency observational surveys ($4$-$11\%$), it appears that the TD simulations overestimate the relative amount of closed magnetic flux that is observed in the heliosphere. While this is certainly possible given the fidelity of our current data-driven simulation capabilities, closed field lines are identified by an instantaneous field-line trace through the model, while time-of-flight and scattering processes dictate the observationally measured suprathermal signal. For closed field lines with an apex well beyond 1AU, it is almost guaranteed that one coronal footpoint will have a much shorter path-length to the in situ measurement point than the other footpoint, and the relative length differences can be extreme for long or highly twisted (or folded) field lines. In this sense one could argue that the measured counterstreaming frequencies are a lower limit on the overall amount of closed flux in the heliosphere. On the other hand the extent to which this loss would take place within the 2 AU TD model domain is debatable and depends strongly on field line length, curvature, and transport processes and needs to be studied further.

Overall our investigation illustrates the importance of including
magnetic flux evolution at the solar surface when
investigating 
magnetic connectivity of the corona and heliosphere. 
In future studies we aim to use this TD modeling paradigm to explore other ways in which combined coronal and heliospheric TD and SS systems inherently differ. This may be especially important for capturing the instantaneous connectivity during solar energetic particle (SEP) events and for identifying solar source regions with ballistic mapping techniques.

The concept of a background steady-state corona and heliosphere can be an incredibly useful construct to understand the physics of solar wind heliospheric structure (e.g. the canonical Parker spiral and ballerina skirt HCS). However, ultimately it is the inherent dynamics of the system---as driven by photospheric flux evolution and associated coronal dynamics---that continually perturb and disrupt this background. Such dynamics can only be captured by truly data-driven MHD modeling, and the relatively good agreement of the TD simulations with strahl statistics and comparatively poor agreement of the SS simulations underscores this point.

%---------------------------------------------------------------------
% FIGURE: FIELD LINE VISUALIZATION
%---------------------------------------------------------------------

\clearpage

\begin{acknowledgments}
 We thank the NASA High-End Computing (HEC) Program through the NASA Advanced Supercomputing Division (NAS) at Ames Research Center for allocations on the Pleiades, Electra, and Aitken supercomputers and the NSF ACCESS program for allocations on the Expanse supercomputer at the San Diego Supercomputer Center (SDSC), which were used to run the simulations. These simulations would not have been possible without the generous support from personnel at both centers, as well as the use of special priority queues. 
This work was supported by the NASA Living With a Star Strategic Capabilities program (80NSSC22K0893), NSF SHINE program (AGS 2501333),  NASA Living With a Star Science program (80NSSC20K0192 and 80NSSC22K1021), and NSF PREEVENTS program (ICER 1854790). We thank Dr.\ Susan T.\ Lepri for a private conversation about strahl measurements.

\end{acknowledgments}

\bibliographystyle{aasjournalv7}
\bibliography{mybib}

\end{document}